\documentclass{svmult}

\usepackage{makeidx}         % allows index generation
\usepackage{graphicx}        % standard LaTeX graphics tool
\usepackage{multicol}        % used for the two-column index
\usepackage[bottom]{footmisc}% places footnotes at page bottom
\makeindex

\begin{document}

\title*{Detailed simulation results for some wealth distribution models in Econophysics}
\titlerunning{Detailed simulation results for some wealth distribution models ...}
\author{K. Bhattacharya$^1$, G. Mukherjee$^{1,2}$, and S. S. Manna$^1$}
\authorrunning{K. Bhattacharya, G. Mukherjee, and S. S. Manna}
\institute{Satyendra Nath Bose National Centre for Basic Sciences \\
           Block-JD, Sector-III, Salt Lake, Kolkata-700098, India \\
\texttt{kunal@bose.res.in, gautamm@bose.res.in, manna@bose.res.in}
\and Bidhan Chandra College, Asansol 713304, Dt. Burdwan, West Bengal, India}
\maketitle

\begin{abstract}
      In this paper we present detailed simulation results on the wealth distribution 
   model with quenched saving propensities. Unlike other wealth distribution models
   where the saving propensities are either zero or constant, this model is not found 
   to be ergodic and self-averaging. The wealth distribution statistics with a single realization 
   of quenched disorder is observed to be significantly different in nature from that of the 
   statistics averaged over a large number of independent quenched configurations. The peculiarities in
   the single realization statistics refuses to vanish irrespective of whatever large 
   sample size is used. This implies that previously observed Pareto law is essentially a
   convolution of the single member distributions.
\end{abstract}

      In a society different members possess different amounts of wealth.
   Individual members often make economic transactions with other members 
   of the society. Therefore in general the wealth of a member fluctuates 
   with time and this is true for all other members of the society as well. 
   Over a reasonably lengthy time interval of observation, which is small 
   compared to the inherent time scales of the economic society this
   situation may be looked upon as a stationary state which implies that 
   statistical properties like the individual wealth distribution, mean 
   wealth, its fluctuation etc. are independent of time.

      More than a century before, Pareto observed that the individual wealth
   $(m)$ distribution in a society is characterized by a power-law tail
   like: $P(m) \sim m^{-(1+\nu)}$ and predicted a value for the constant
   $\nu \approx 1$, known as the Pareto exponent \cite {Pareto}. Very recently, i.e., over the
   last few years, the wealth distribution in a society has attracted renewed
   interests in the context of the study of {\it Econophysics} and various models
   have been proposed and studied. A number of analyses have also been done
   on the real-world wealth distribution data in different countries \cite 
   {realdatag,realdataln,Sitabhra}. All these recent 
   data indeed show that Pareto like power-law tails do exist in the wealth 
   distributions in the large wealth regime but with different values of
   the Pareto exponent ranging from $\nu=1$ to 3. It has also been observed
   that only a small fraction of very rich members actually contribute to the
   Pareto behavior whereas the middle and the low wealth individuals follow either
   exponential or log-normal distributions.

      In this paper we report our detailed simulation results on the three recent
   models of wealth distribution. The three models are:
   (i) the model of  Dr\u{a}gulescu and Yakovenko (DY) \cite {DY} which gives
   an exponential decay of the wealth distribution, (ii) the model of Chakraborti and
   Chakrabarti (CC) \cite {CC} with a fixed saving propensity giving a
   Gamma function for the wealth distribution and (iii) the model of Chatterjee,
   Chakrabarti and Manna (CCM) \cite {CCM} with a distribution of quenched
   individual saving propensities giving a Pareto law for the wealth distribution.

%---------------------------------------------------------------------------
\begin{figure}[top]
\begin{center}
\includegraphics[width=12.0cm]{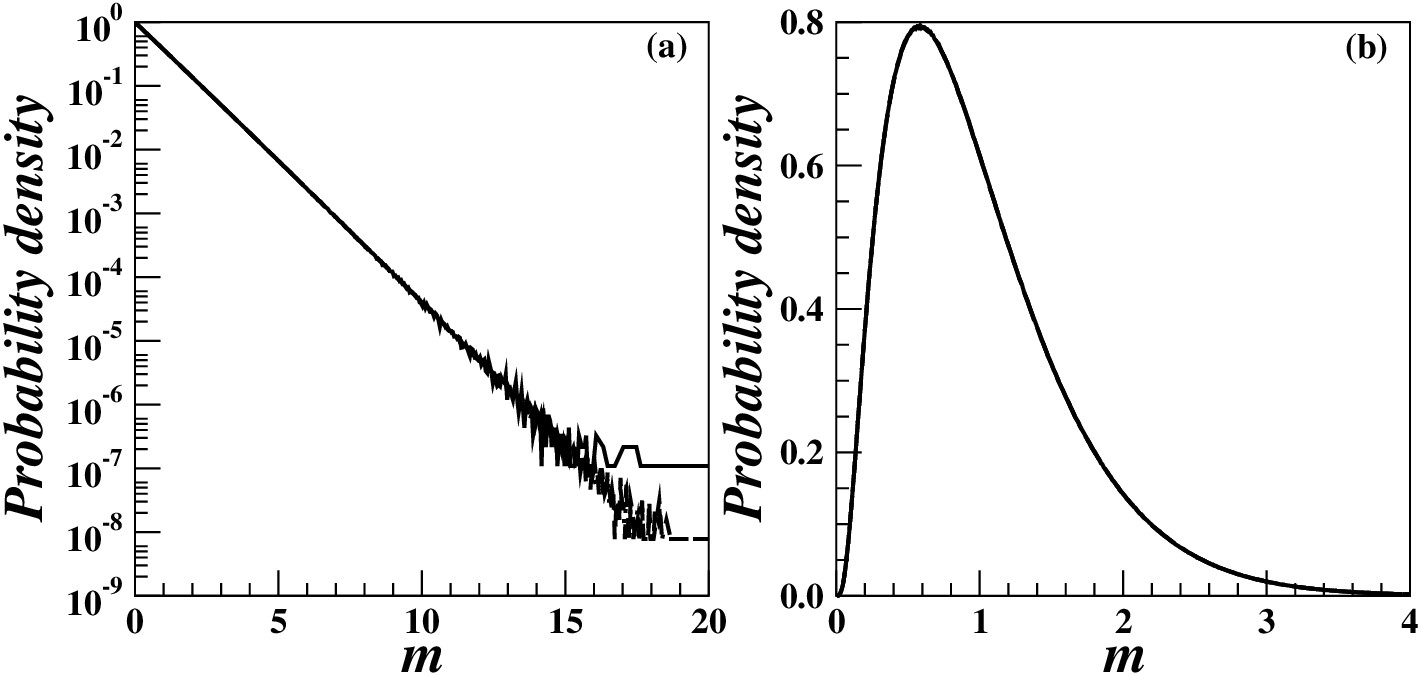}
\end{center}
\caption{
The three probability densities of wealth distribution, namely ${\rm Prob_1}(m)$
(solid line), ${\rm Prob_2}(m)$ (dashed line) and ${\rm Prob}(m)$ (dot-dashed 
line) are plotted with wealth $m$ for $N$ = 256 in (a) for the DY model and in 
(b) for the CC model for $\lambda$ = 0.35. The excellent overlapping of all 
three curves indicate that both the DY and CC models are ergodic as well as 
self averaging.
}
\end{figure}
%---------------------------------------------------------------------------

      All these three models have some common features. The society consists
   of a group of $N$ individuals, each has a wealth $m_i(t), i=1,N$. The wealth
   distribution $\{m_i(t)\}$ dynamically evolves with time following the pairwise
   conservative money shuffling method of economic transactions. Randomly selected pairs of individuals
   make economic transactions one after another in a time sequence and thus the
   wealth distribution changes with time. For example, let two randomly selected
   individuals $i$ and $j$, $(i \ne j)$ have wealths $m_i$ and $m_j$. They make
   transactions by a random bi-partitioning of their total wealth $m_i+m_j$ and then
   receiving one part each randomly:
\begin {eqnarray}
& m_i(t+1)={\epsilon(t)}(m_i(t)+m_j(t)) \nonumber \\
& m_j(t+1)={(1-\epsilon(t))}(m_i(t)+m_j(t)).
\end {eqnarray}
   Here time $t$ is simply the number of transactions and $\epsilon(t)$ is the
   $t$-th random fraction with uniform distribution drawn for the $t$-th transaction.

      In all three models the system dynamically evolves to a 
   stationary state which is characterized by a time independent probability
   distribution ${\rm Prob}(m)$ of wealths irrespective of the details of the initial
   distribution of wealths to start with. Typically in all our simulations a
   fixed amount of wealth is assigned to all members of the society, i.e.
   ${\rm Prob}(m,t=0) = \delta(m-\langle m \rangle)$. The model described so far is precisely the
   DY model in \cite {DY}. The stationary state wealth distribution for
   this model is \cite {DY,DY1,Das:2003}:
\begin {equation}
{\rm Prob}(m) = \frac{1}{\langle m \rangle} \exp (-m/\langle m \rangle).
\end {equation}
   Typically $\langle m \rangle$ is chosen to be unity without any loss of generality.

%---------------------------------------------------------------------------
\begin{figure}[top]
\begin{center}
\includegraphics[width=12.0cm]{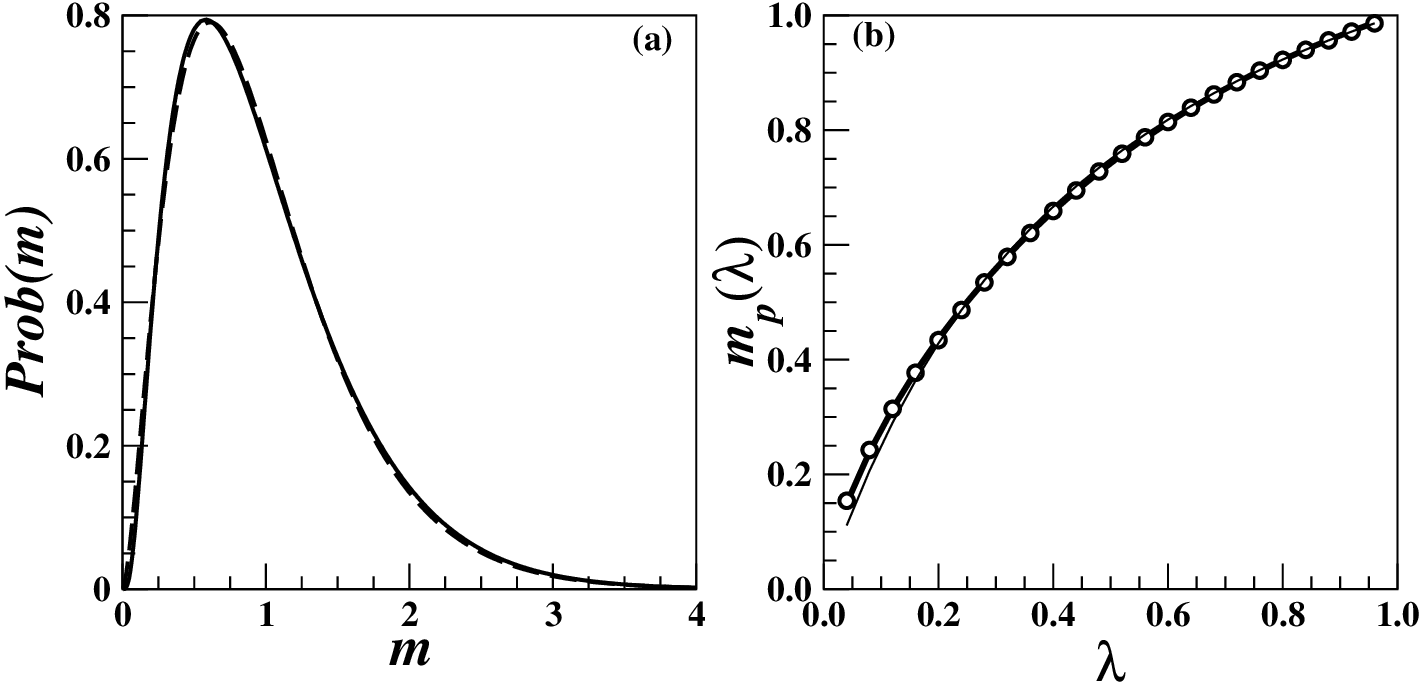}
\end{center}
\caption{
For the CC model with $N$ = 256 and $\lambda$ = 0.35 these plots show the
functional fits of the wealth distribution in (a) and the variation of the 
most probable wealth $m_p(\lambda)$ in (b). In (a) the simulation data of
${\rm Prob}(m)$ is shown by the solid black line where as the fitted Gamma
function of Eqn. (5) is shown by the dashed line. In (b) the $m_p(\lambda)$
data for 24 different $\lambda$ values denoted by circles is fitted to the 
Gamma function given in Eqn. (6) (solid line).
The thin line is a comparison with the $m_p(\lambda)$ values obtained from the analytical expression
of $a(\lambda)$ and $b(\lambda)$ in \cite {Patriarca:2004}.
}
\end{figure}
%---------------------------------------------------------------------------

      A fixed saving propensity is introduced in the CC model \cite {CC}. During the
   economic transaction each member saves a constant $\lambda$ fraction of his wealth.
   The total sum of the remaining wealths of both the traders is then randomly
   partitioned and obtained by the individual members randomly as follows:
%\begin {center}
\begin {eqnarray}
& m_i(t+1)= \lambda m_i(t) + {\epsilon(t)}(1-\lambda)(m_i(t)+m_j(t)) \nonumber \\
& m_j(t+1)= \lambda m_j(t) + {(1-\epsilon(t))}(1-\lambda)(m_i(t)+m_j(t)).
\end {eqnarray}
%\end {center}
   The stationary state wealth distribution is an asymmetric distribution with a
   single peak. The distribution vanishes at $m=0$ as well as for large $m$ values.
   The most probable wealth $m_p(\lambda)$ increases monotonically with $\lambda$ and the
   distribution tends to
   the delta function again in the limit of $\lambda \to 1$ irrespective of the
   initial distribution of wealth.

      In the third CCM model different members have their own fixed individual saving
   propensities and therefore the set of $\{\lambda_i, i=1,N\}$ is a quenched variable.
   Economic transactions therefore take place following these equations:
%\begin {center}
\begin {eqnarray}
& m_i(t+1)=\lambda_im_i(t) +   \epsilon(t) [(1-\lambda_i)m_i(t)+(1-\lambda_j)m_j(t)] \nonumber \\
& m_j(t+1)=\lambda_jm_j(t) +(1-\epsilon(t))[(1-\lambda_i)m_i(t)+(1-\lambda_j)m_j(t)]
\end {eqnarray}
%\end {center}
   where $\lambda_i$ and $\lambda_j$ are the saving propensities of the members $i$ and
   $j$. The stationary state wealth distribution shows a power law decay
   with a value of the Pareto exponent $\nu \approx 1$ \cite {CCM}.

      In this paper we present the detailed numerical evidence to argue
   that while the first two models are ergodic and
   self-averaging, the third model is not. This makes the third model
   difficult to study numerically.

%---------------------------------------------------------------------------
\begin{figure}[top]
\begin{center}
\includegraphics[width=10.5cm]{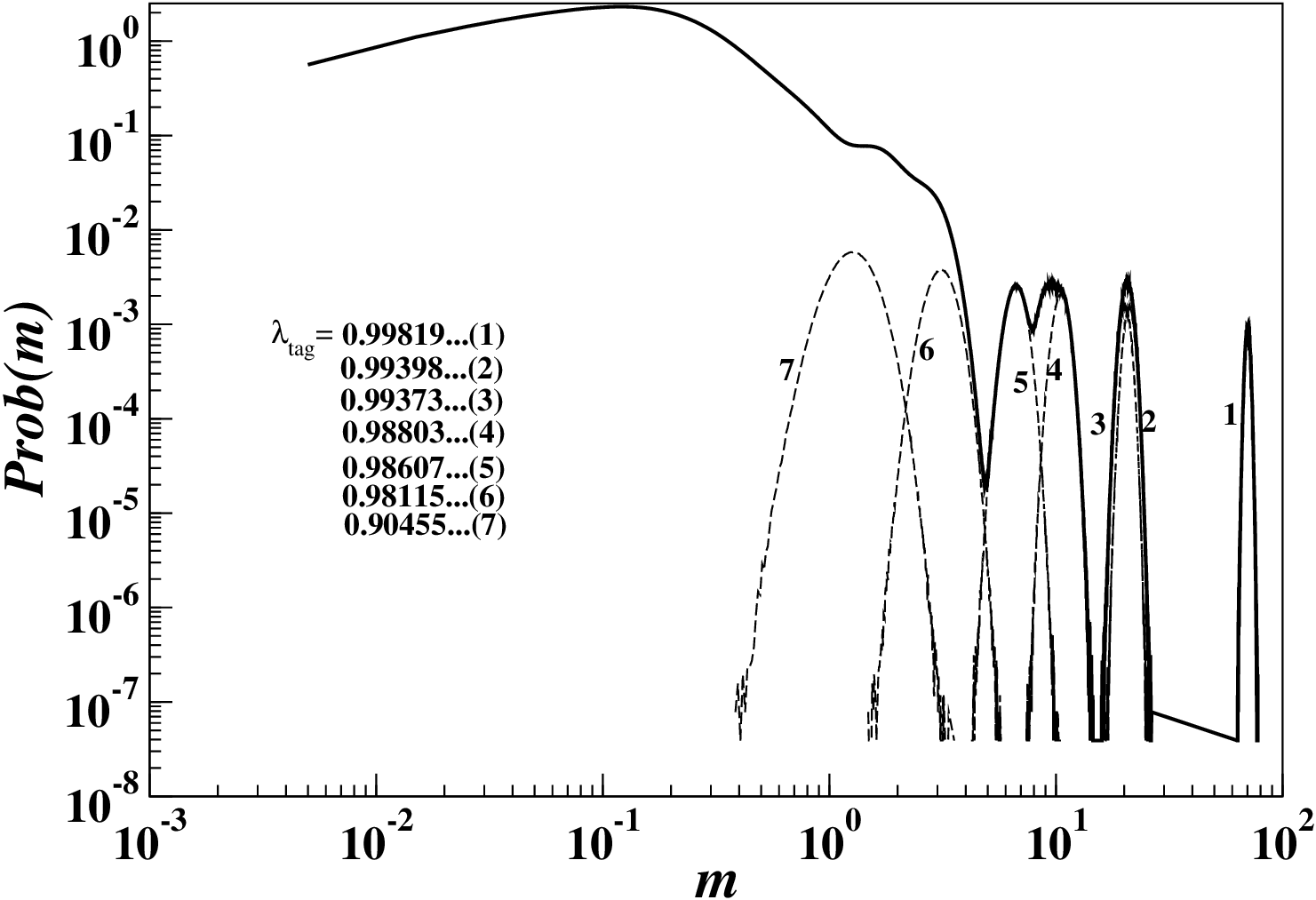}
\end{center}
\caption{
The wealth distribution ${\rm Prob}(m)$ in the stationary state for the CCM model for a 
single initial configuration of saving propensities $\{\lambda_i\}$ with 
$N$=256 is shown by the solid line. Also the wealth distributions of the
individual members with seven different tagged values of $\lambda_{tag}$
are also plotted on the same curve with dashed lines. This shows that the
averaged (over all members) distribution ${\rm Prob}(m)$ is the convolution
of wealth distributions of all individual members.
}
\end{figure}
%---------------------------------------------------------------------------

      We simulated DY model with $N=256, 512$ and $1024$. Starting from an initial
   equal wealth distribution ${\rm Prob}(m)=\delta(m-1)$ we skipped some transactions
   corresponding to a relaxation time $t_{\times}$ to reach the stationary state. Typically
   $t_{\times} \propto N$. In the stationary state we calculated the three different probability
   distributions, namely: (i) the wealth distribution ${\rm Prob_1}(m)$ of an arbitrarily selected tagged member
   (ii) the overall wealth distribution ${\rm Prob_2}(m)$ (averaged over all members of the society)
   on a long single run (single initial configuration, single sequence of random numbers) and
   (iii) the overall wealth distribution ${\rm Prob}(m)$ averaged over many initial configurations.
   In Fig. 1(a) we show all three plots for $N=256$ and observe that these three plots
   overlap excellent, i.e., these distributions are same. This implies that
   the DY model is ergodic as well as self-averaging.

%---------------------------------------------------------------------------
\begin{figure}[top]
\begin{center}
\includegraphics[width=12.0cm]{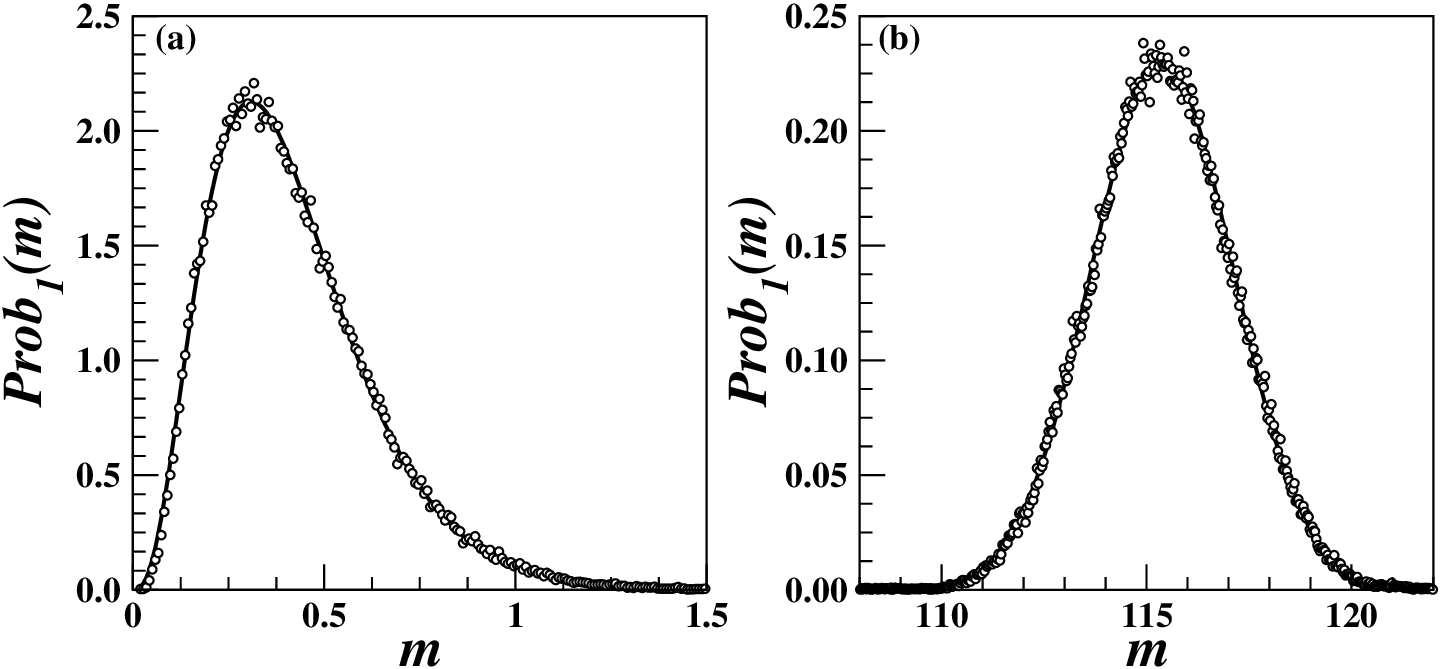}
\end{center}
\caption{
The individual member's wealth distribution in the CCM model. A member is tagged
with a fixed saving propensity $\lambda_{tag}$=0.05 in (a) and 0.999 in (b) for $N$=256.
In the stationary state the distribution ${\rm Prob_1}(m)$ is asymmetric in (a)
and is fitted to a Gamma function. However for very large
$\lambda$ the distribution in (b) is symmetric and fits very nicely to a
Gaussian distribution.
}
\end{figure}
%---------------------------------------------------------------------------

      Similar calculations are done for the CC model as well (Fig. 1(b)). We see a
   similar collapse of the data for the same three probability distributions. This lead us to
   conclude again that the CC model is also ergodic and self-averaging. Further we fit in Fig. 2(a) the
   CC model distribution ${\rm Prob}(m)$ using a Gamma function as cited in \cite {Patriarca:2004}
   as:
\begin {equation}
{\rm Prob}(m) \sim m^{a(\lambda)} \exp(-b(\lambda)m)
\end {equation}
   which gives excellent non-linear fits by ${\it xmgrace}$ to all values of $\lambda$
   in the range between say 0.1 to 0.9. Once fitting is done the most-probable wealth
   is estimated by the relation: $m_p(\lambda)=a(\lambda)/b(\lambda)$ using the values
   of fitted parameters $a(\lambda)$ and $b(\lambda)$. Functional dependences of
   $a$ and $b$ on $\lambda$ are also predicted in \cite {Patriarca:2004}. We plot
   $m_p(\lambda)$ so obtained with $\lambda$ for 24 different values of $\lambda$ in 
   Fig. 2(b). We observe that these data points fit very well to another Gamma function as:
\begin {equation}
m_p(\lambda) = A \lambda^{\alpha} \exp(-{\beta}\lambda).
\end {equation}
   The values of $A \approx 1.46$, $\alpha \approx 0.703$ and $\beta \approx 0.377$ are estimated
   for $N$ = 256, 512 and 1024 and
   we observe a concurrence of these values up to three decimal places for the
   three different system sizes. While $m_p(0)=0$ from Eqn. (6) is 
   consistent, $m_p(1)=1$ implies $A=\exp(\beta)$ is also consistent with estimated values
   of $A$ and $\beta$. Following \cite {Patriarca:2004} we plotted $m_p(\lambda)= 3\lambda/(1+2\lambda)$
   in Fig. 2(b) for the same values of $\lambda$
   and observe that these values deviate from our points for the small values of $\lambda$.

      However, for the CCM model many inherent structures are observed. 
   We argue that this model is neither ergodic nor self-averaging.
   For a society of $N=256$ members a set of quenched individual saving propensities
   $\{0 \le \lambda_i\ < 1, i=1,N\}$ are assigned drawing these numbers from an independent and
   identical distribution of random numbers. The system then starts evolving with
   random pairwise conservative exchange rules cited in Eqn. (4). First we reproduced
   the ${\rm Prob}(m)$ vs. $m$ curve given in \cite {CCM} by averaging the
   wealth distribution over 500 uncorrelated initial configurations. The data
   looked very similar to that given in \cite {CCM} and the Pareto exponent $\nu$ is
   found to be very close to 1. 

%---------------------------------------------------------------------------
\begin{figure}[top]
\begin{center}
\includegraphics[width=12.0cm]{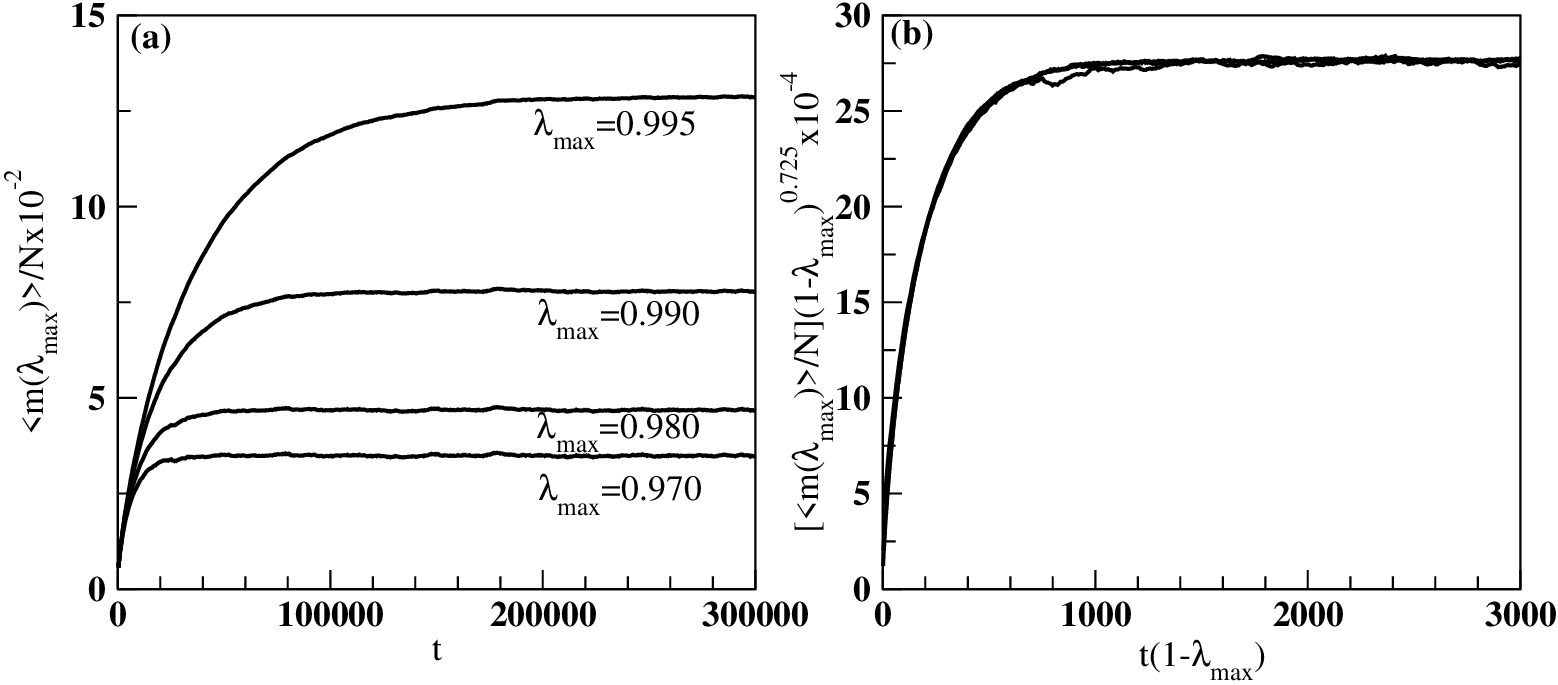}
\end{center}
\caption{
(a) The mean wealth of a tagged member who has the maximal saving propensity is
plotted as a function of time for four different values of $\lambda_{max}$.
In (b) this data is scaled to obtain the data collapse.
}
\end{figure}
%---------------------------------------------------------------------------

      Next we plot the same data for a single quenched configuration
   of saving propensities as shown in Fig. 3. It is observed that the 
   wealth distribution plotted by the continuous solid line is far from being
   a nice power law as observed in \cite {CCM} for the configuration averaged distribution.
   This curve in Fig. 3 has many humps, especially
   in the large wealth limit. To explain this we made further simulations
   by keeping track of the wealth distributions of the individual members.
   We see that the individual wealth distributions are significantly different from being 
   power laws, they have single peaks as shown in Fig. 4. For small values of
   $\lambda$, the ${\rm Prob_1}(m)$ distribution is asymmetric and has the form of a 
   Gamma function similar to what is already observed for the CC model (Fig. 4(a)). On the 
   other hand as $\lambda \to 1$ the variation becomes more and more symmetric 
   which finally attains a simple Gaussian function (Fig. 4(b)). The reason
   is for small $\lambda$ the individual wealth distribution does feel the 
   presence of the infinite wall at $m=0$ since no debt is allowed in this 
   model, where as for $\lambda \to 1$ no such wall is present and consequently the distribution
   becomes symmetric. This implies that the wealth possessed by an individual
   varies within a limited region around an average value and certainly the
   corresponding phase trajectory does not explore the whole phase space.
   Therefore we conclude that the CCM model is not ergodic. 

      Seven individual wealth distributions have been plotted in Fig. 3. 
   corresponding to six top most $\lambda$ values and one with somewhat
   smaller value. We see that top parts of these ${\rm Prob_1}(m)$ distributions almost
   overlap with the ${\rm Prob_2}(m)$ distribution. This shows that ${\rm Prob_2}(m)$
   distribution is truly a superposition of $N$ ${\rm Prob_1}(m)$ distributions.
   In the limit of $\lambda \to 1$, large gaps are observed in the ${\rm Prob_2}(m)$ 
   distribution due
   to slight differences in the $\lambda$ values of the corresponding
   individuals. These gaps remain there no matter whatever large sample
   size is used for the ${\rm Prob_2}(m)$ distribution.

%---------------------------------------------------------------------------
\begin{figure}[top]
\begin{center}
\includegraphics[width=12.0cm]{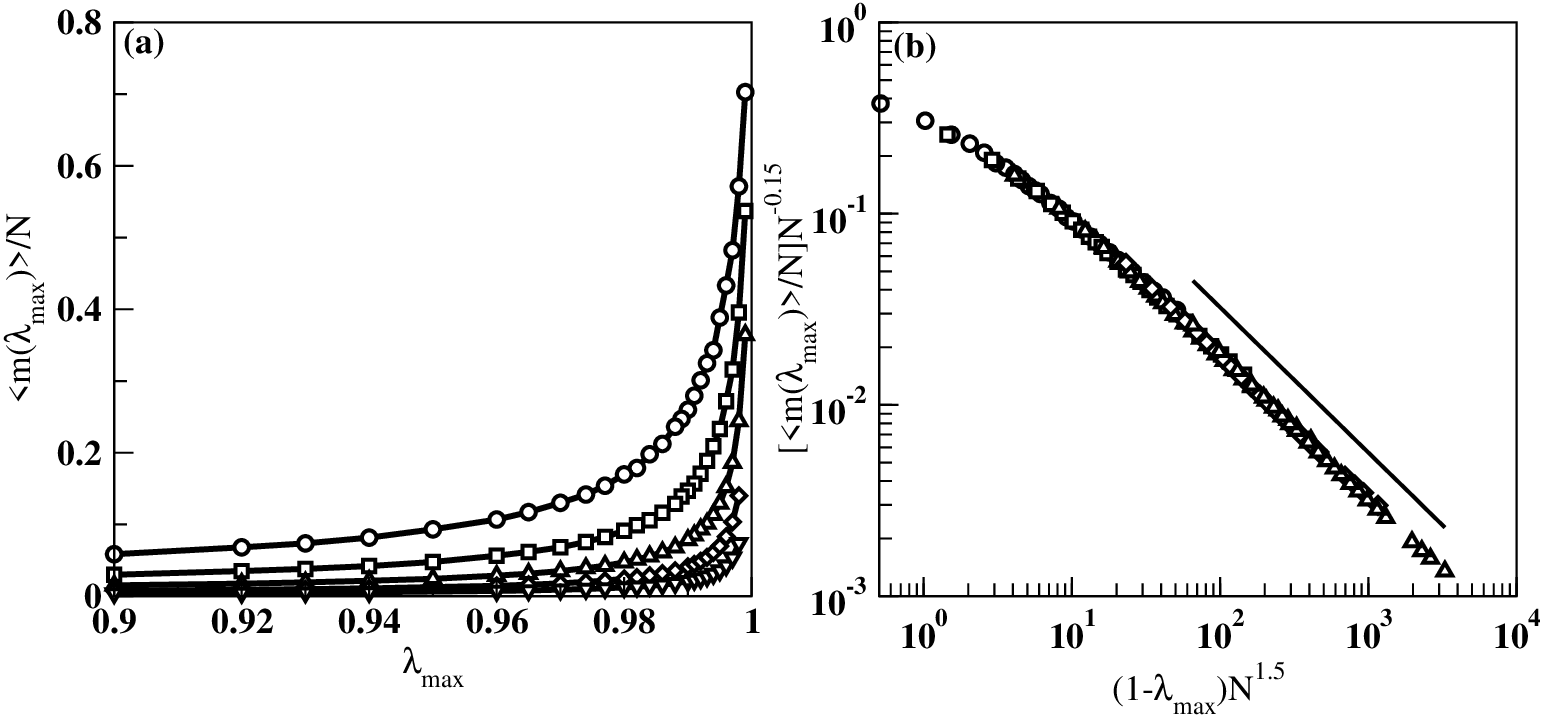}
\end{center}
\caption{
In the stationary state the mean value of the wealth of the member with 
maximum saving propensity $\lambda_{max}$ is plotted with $\lambda_{max}$.
This value diverges as $\lambda_{max} \to 1$ for $N$ = 64 (circle), 128 (square), 256
(triangle up), 512 (diamond) and 1024 (triangle down). 
(b) This data is scaled to obtain a data collapse of the three different sizes.
}
\end{figure}
%---------------------------------------------------------------------------

      We further argue that even the configuration averaging may be difficult
   due to very slow relaxation modes present in the system. To demonstrate this point we consider
   the CCM model where the maximal saving propensity $\lambda_{max}$ is
   continuously tuned. The $N$-th member is assigned $\lambda_{max}$ and
   all other members are assigned values $\{0 \le \lambda_i < \lambda_{max}, i=1,N-1\}$.
   The average wealth $\langle m(\lambda_{max}) \rangle/N$ of the $N$-th member
   is estimated at different times for $N$ = 256 and they are plotted in Fig. 5(a) for
   four different values of $\lambda_{max}$. It is seen that as $\lambda_{max} \to 1$ 
   it takes increasingly longer relaxation times to reach the stationary state and
   the saturation value of the mean wealth in the stationary state also increases very rapidly. 
   In Fig. 5(b) we made a scaling of these plots like
\begin {equation}
[\langle m(\lambda_{max}) \rangle/N](1-\lambda_{max})^{0.725} \sim {\cal G}[t(1-\lambda_{max})].
\end {equation}
   This implies that the stationary state of the member with maximal
   saving propensity is reached after a relaxation time $t_{\times}$ given by
\begin {equation}
t_{\times} \propto (1-\lambda_{max})^{-1}.
\end {equation}
   Therefore we conclude that in CCM the maximal $\lambda$ member 
   takes the longest time to reach the stationary state
   where as rest of the members reach their individual stationary
   states earlier.

      This observation poses a difficulty in the simulation of the CCM
   model. Since this is a problem of quenched disorder it is necessary
   that the observables should be averaged over many independent realizations
   of uncorrelated disorders. Starting from an arbitrary initial distribution
   of $m_i$ values one generally skips the relaxation time $t_{\times}$ 
   to reach the stationary state and then
   collect the data. In the CCM model the $0 \le \lambda_i < 1$ is used.
   Therefore if $M$ different quenched disorders are used for averaging it means the
   maximal of all $M \times N$ $\lambda$ values is around $1-1/(MN)$. From Eqn. (8) this
   implies that the slowest relaxation time grows proportional to $MN$. Therefore the main
   message is more accurate simulation one intends to do by increasing the
   number of quenched configurations, larger relaxation time $t_{\times}$ it has to skipp for each
   quenched configuration to ensure that it had really reached the stationary state.
 
      Next, we calculate the variation of the mean wealth $\langle m(\lambda_{\max}) \rangle / N$
   of the maximally tagged member in the stationary state as a function of $\lambda_{max}$ and
   for the different values of $N$. In Fig. 6(a) we plot this variation for 
   $N$ = 64, 128, 256, 512 and 1024 with different symbols. It is observed that larger
   the value of $N$ the $\langle m(\lambda_{max}) \rangle/N$ is closer to zero for all values of
   $\lambda_{max}$ except for those which are very close to 1. For
   $\lambda_{max} \to 1$ the mean wealth increases very sharply to achieve
   the condensation limit of $\langle m(\lambda_{max}=1) \rangle / N = 1$.

      It is also observed that the divergence of the mean wealth near 
   $\lambda_{max}=1$ is associated with a critical exponent. In Fig. 6(b)
   we plot the same mean wealth with the deviation $(1-\lambda_{max})$
   from 1 on a double logarithmic scale and observe power law variations.
   A scaling of these plots is done corresponding to a data collapse like:
\begin {equation}
[\langle m(\lambda_{max}) \rangle/N]N^{-0.15} \sim {\cal F}[(1-\lambda_{max})N^{1.5}].
\end {equation}
   Different symbols representing the data for the same five system sizes
   fall on the same curve which has a slope around 0.76. The scaling function
   ${\cal F}[x] \to x^{-\delta}$ as $x \to 0$ with $\delta \approx 0.76$. This 
   means $\langle m(\lambda_{max}) \rangle N^{-1.15} \sim (1-\lambda_{max})^{-0.76}N^{-1.14}$
   or $\langle m(\lambda_{max}) \rangle \sim (1-\lambda_{max})^{-0.76}N^{0.01}$. 
   Since for a society of $N$ traders $(1-\lambda_{max}) \sim 1/N$
   this implies 
\begin {equation}
\langle m(\lambda_{max}) \rangle \sim N^{0.77}.
\end {equation}
   This result is therefore different from the claim that $\langle m(\lambda_{max}) \rangle \sim N$
   \cite {CCM}.

      To summarize, we have revisited the three recent models of wealth distribution
   in Econophysics. Detailed numerical analysis yields that while the DY and CC models
   are ergodic and self-averaging, the CCM model with quenched saving propensities
   does not seem to be so. In CCM existence of slow modes proportional to the total sample
   size makes the numerical analysis difficult. 

      All of us thank B. K. Chakrabarti and S. Yarlagadda and A. Chatterjee for their
   very nice hospitality in the {\bf ECONOPHYS - KOLKATA I} meeting. GM thankfully
   acknowledged facilities at S. N. Bose National Centre for Basic Sciences in the
   FIP programme.

\end {document}